\documentclass[twocolumn]{revtex4}
\usepackage{graphicx}
\usepackage{dcolumn}
\usepackage{longtable}

\begin{document}

\title{Regularities and symmetries of subsets of collective $0^+$ states}

\author{Dennis Bonatsos$^1$, E.A. McCutchan$^2$, R.F. Casten$^3$, R.J. Casperson$^3$, V. Werner$^3$, and E. Williams$^3$}

\affiliation{$^1$ Institute of Nuclear Physics, N.C.S.R. ``Demokritos'', GR-15310 Aghia Paraskevi, Attiki, Greece}
\affiliation{$^2$ Physics Division, Argonne National Laboratory, Argonne, Illinois 60439, USA}
\affiliation{$^3$ Wright Nuclear Structure Laboratory, Yale University, New Haven, CT 06520, USA}

\begin{abstract}
The energies of subsets of excited $0^+$ states in geometric collective models are investigated and found to exhibit intriguing regularities.  In models with an infinite square well potential, it is found that a single formula, dependent on only the number of dimensions, describes a subset of $0^+$ states. The same behavior of a subset of $0^+$ states is seen in the large boson number limit 
of the Interacting Boson Approximation (IBA) model near the critical point of a first order phase transition, in contrast to the fact that  
these $0^+$ state energies exhibit a harmonic behavior in all three limiting symmetries of the IBA. Finally, the observed regularities in $0^+$ energies are analyzed in terms of the underlying group theoretical framework of the different models. 

\end{abstract}

\maketitle

\section{Introduction} % 1 

One of the overarching themes of the science of complex many-body systems is to understand the remarkable regularities they often exhibit and try to relate these to underlying symmetries of the system. In nuclei, this challenge is approached through the use of geometric and algebraic models that describe collective behavior of the nuclear system.  There are a large number and variety of such models, each with seemingly unique properties and predictions.  Nevertheless, careful analysis often shows relations among such models that have escaped notice and therefore leads to a better understanding of their mutual interrelationships and often, to experimental tests and constraints on their applicability. 

In the present work, we will focus on the properties of subsets of $0^+$ states in nuclei. In general, $0^+$ states are of fundamental importance since they are easily observed experimentally in reactions such as few nucleon transfer \cite{Oothoudt,Lesher,des} or beta decay \cite{Asai}.
Although not all $0^+$ states are collective in nature, they are always intrinsic excitations of the ground state condensate, and are free of some of the complications (such as centrifugal effects) present in other states.  In the present work, we focus on the properties of a subset of collective $0^+$ states.  We will show that broad classes of seemingly diverse models actually yield identical predictions for that subset of $0^+$ states. 
One upshot of this analysis will be the development of extremely simple, analytic eigenvalue expressions which, in one case, depend only on the dimensionality of the system and, in another, turn out to transcend the symmetry structure.  

\begin{figure}
\includegraphics[height=100mm]{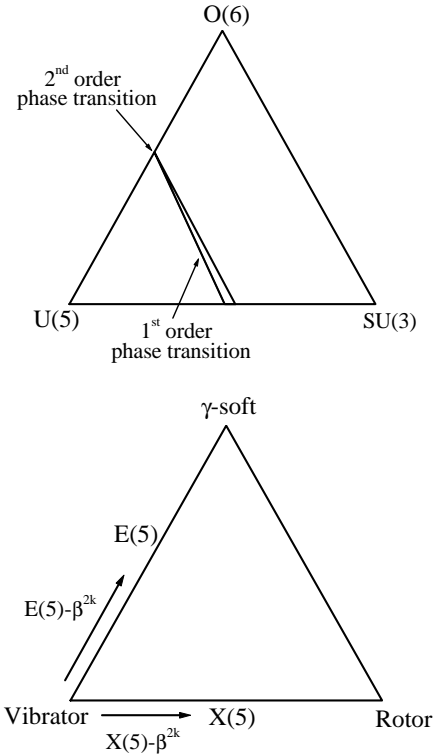}
\caption{(Top) Symmetry triangle with the three IBA dynamical symmetries placed at the vertices. The phase transition region is indicated by the slanted lines. 
(Bottom) Similar triangle in a geometrical framework. The critical point symmetries E(5) and X(5) are located close to the phase transition region of the IBA.  The models E(5)-$\beta^{2k}$ and X(5)-$\beta^{2k}$ span structures between a vibrator and the critical point symmetries.}
\end{figure}

Our approach primarily exploits two classes of models, namely the Interacting Boson Approximation (IBA) model and geometric descriptions of nuclei at critical points of quantum phase transitions in their equilibrium structure. Therefore, we start by briefly recalling pertinent features of these models.  
The Interacting Boson Approximation model~\cite{iba} describes collective structure in terms of bosons of angular momentum zero ($s$-bosons) and two ($d$-bosons) in the framework of an overall U(6) symmetry.  Emerging from the U(6) symmetry group structure are three dynamical symmetries which have long been benchmark paradigms of structure : U(5), which gives vibrational structure characteristic of spherical nuclei, SU(3), which describes axially symmetric deformed rotors, and O(6), which pertains to deformed nuclei that are soft with respect to axial asymmetry ($\gamma$-soft). 
Shape/phase transitions in atomic nuclei were discussed~\cite{gilmore} many years ago in the classical analog~\cite{GK,DSI} of the IBA.

To visualize these limiting symmetries and the transitions between them, it is common to place them at the corners of a symmetry triangle~\cite{ricktri}, as shown in Fig. 1(top).  In the IBA framework, a point of first order phase transition occurs between U(5) and SU(3), while a point of second order phase transition occurs between U(5) and O(6).  The triangle is divided into two regions, spherical and deformed, by a narrow shape coexistence region~\cite{IZC} extending around the line of first order phase transition connecting the two points mentioned above. 
In the classical limit of the IBA, obtained through use of the intrinsic state formalism \cite{GK,DSI},
one can use \cite{Jolie89} Landau theory to delineate a similar phase transitional behavior. 

\begin{figure}
\includegraphics[height=60mm]{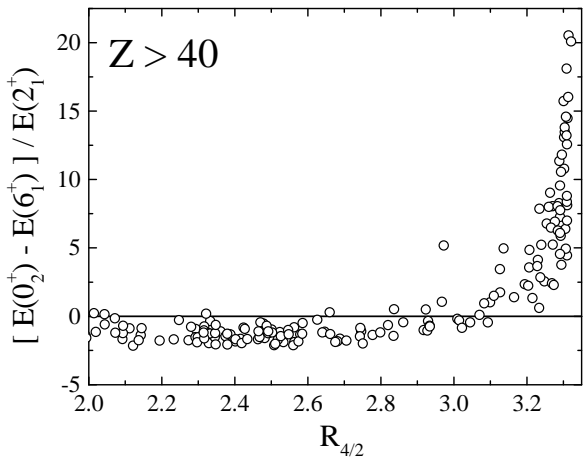}
\caption{Experimental [$E$($0_2^+$)-$E$($6_1^+$)]/$E$($2_1^+$) values plotted as a function of the $R_{4/2}$ $\equiv$ $E$($4_1^+$)/$E$($2_1^+)$ ratio for all even-even nuclei with $Z>40$.
Data from Ref.~\cite{nds}.}
\end{figure} 

More recently, phase transitions have been investigated in a geometrical framework. The critical point symmetries E(5)~\cite{IacE5} and X(5)~\cite{IacX5} have been developed to describe phase/shape transitions between vibrational to $\gamma$-soft and vibrational to axially symmetric deformed, respectively, using special solutions of the Bohr Hamiltonian~\cite{bohr}. These solutions are analytic and parameter free, except for scale. In Fig. 1(bottom), we indicate their position close to the critical point of a phase transition in a symmetry triangle for geometrical models.  The concept of critical point symmetries (CPS) is supported by the observation of nuclei exhibiting such properties~\cite{CZE5,CZX5,Kruecken,tonev,frank,JPGreview}.  Their success has spawned the development of numerous additional geometrical models, several of which offer analytic solutions and cover a wider range of structures both before and after the phase transitional point.  

Links between the geometrical approach and the IBA formalism have also found renewed interest.  A powerful method for solving the Bohr Hamiltonian numerically has recently been developed~\cite{Rowe735}, leading to an algebraic collective model~\cite{Rowe753}. Examples of the use of this method 
have recently been presented \cite{RWC}. 
The relationship between the algebraic collective model and the different limiting symmetries of the IBA has been studied in Refs.~\cite{Thiamova1,Thiamova2}.  The connection between geometrical models spanning structure near E(5) and the IBA has also been investigated~\cite{ramos}. 

A key prediction of the CPS involves the energy of the first excited $0^+$ state. 
The nature of low-lying $0^+$ states is critical to understanding the structure of nuclei and changes in structure~\cite{des,mapping,cejnar}. While there is some debate as to the nature~\cite{pgarrett} of low-lying $0^+$ states, the fact that the energies of $0^+$ states evolve rather smoothly as a function of changing structure cannot be ignored. This has been pointed out previously in Ref.~\cite{wchou}.  We illustrate this in a similar way in Fig. 2, plotting the relationship between a level of the ground state band, $E$($6_1^+$), and the first excited $0^+$ state, $E$($0_2^+$), as a function of $R_{4/2}$ $\equiv$ $E$($4_1^+$)/$E$($2_1^+$)
for all even-even nuclei with $Z$ $>$ 40.  Despite the enormous range of structures encompassed in the plot, an overall compact trajectory emerges.

The primary purpose of the present work is to investigate the energies of $0^+$ states in a wide range of models. In some cases the analysis applies to all collective $0^+$ states, in others to classes of $0^+$ states that act as bandheads for major families of states.  For a broad class of models, we will discuss some remarkable regularities, develop analytic expressions for the eigenvalues of these states that depend only on the dimensionality of the system, relate these results to more general models, and discuss the group theoretical properties underlying these regularities.  
We begin within the framework of the different solutions of the Bohr Hamiltonian.  We then do a similar analysis within the framework of the IBA and then finally investigate the links between these two different approaches.  Some of this material has been previously summarized~\cite{map,letter}.

\section{$0^+$ states in solutions of the Bohr Hamiltonian}  % 2 

Numerous models are emerging which provide a reasonable description of nuclei using an infinite square well potential.  E(5) and X(5) are special solutions of the Bohr Hamiltonian describing collective nuclear properties in terms of the shape variables $\beta$ and $\gamma$. Both take the potential in $\beta$ as an infinite square well but use different potentials in $\gamma$;
X(5) uses a harmonic oscillator potential in $\gamma$ which has a minimum at $\gamma=0^{\circ}$ whereas E(5) takes a potential independent of $\gamma$.  Additional solutions which make use of infinite square well potentials in $\beta$ include Z(5)~\cite{Z5}, which uses a harmonic oscillator potential in $\gamma$ with a minimum at $\gamma$ = 30$^{\circ}$, Z(4)~\cite{Z4} where $\gamma$ is frozen to 30$^{\circ}$, and X(3)~\cite{X3} where $\gamma$ is fixed at 0$^{\circ}$.

In each of the infinite square well solutions, the energy eigenvalues are proportional to the squares of roots of the Bessel functions, $J_{\nu}(z)$, where the order $\nu$ is different for each solution.  The orders of the Bessel functions obtained in the E(5), X(5), Z(5), Z(4), and X(3) models are summarized in Table I, along with the dimension, $D$, of each model and the value of $\nu$ for $J^{\pi}=0^+$ states.  The dimension effectively refers to the number of degrees of freedom of the model.  For example, the five dimensional models are described by $\beta$, $\gamma$ and the three Euler angles. 

In the X(3) and Z(4) models, we consider all excited $0^+$ states. In X(5) and Z(5), the solutions are obtained through an approximate separation of the $\beta$ and $\gamma$ degrees of freedom.  We consider those $0^+$ states arising from the $\beta$ solution, as they are directly related to the infinite square well potential.  In the E(5) solution, we consider those $0^+$ states with $\tau$ = 0, that is, those $0^+$ states which correspond to base states on which major families of levels are built.  

\begin{table}
\caption{Order $\nu$, dimension, $D$, of the model space and $\nu$ for $J^{\pi}=0^+$ states in the geometrical models E(5), X(5), Z(5), Z(4), and X(3). $J$ is the spin of the level, $\tau=J/2$ (in the ground state band), and $n_{w}$ is the wobbling quantum number~\cite{BM}, which is zero for $0^+$ states.}
\renewcommand{\arraystretch}{2.0}
\begin{tabular}{lccc}
\hline

Model & $\nu$ & $\;$$\;$$\;$$\;$ D $\;$$\;$$\;$$\;$ & $\nu$($J$=$0^+$) \\

\hline

E(5) & $\tau$ + $\frac{3}{2}$ & 5 & $\frac{3}{2}$ \\

\hline

X(5) & $\sqrt{\frac{J(J+1)}{3} + \frac{9}{4}}$ & 5 &  $\frac{3}{2}$ \\

\hline 

Z(5) & $\frac{\sqrt{J(J+4)+3n_w(2J-n_w)+9}}{2}$ & 5 &  $\frac{3}{2}$ \\

\hline

Z(4) & $\frac{\sqrt{J(J+4)+3n_w(2J-n_w)+4}}{2}$ & 4 &  1 \\

\hline

X(3) & $\sqrt{\frac{J(J+1)}{3}+\frac{1}{4}}$ & 3 & $\frac{1}{2}$ \\

\hline

\end{tabular}
\end{table}

Traditionally, the excitation energy of the first excited $2^+$ state is used to set the overall scale for these models. However, in some cases, using a different normalization can reveal physics not otherwise very evident. In particular, one can sometimes see relations among states of the same angular momentum by normalizing to the first excited state of that spin.  Hence, here, we scale to the energy of the first excited $0^+$ state, $0_2^+$.  It turns out that this approach allows the relative energies of these excited $0^+$ states to be well described by simple formulas. For the $0^+$ states we use the usual notation, $0_m^+$, where $m=1$ corresponds to the ground state, $m=2$ denotes 
the first excited $0^+$ state, and so on, within each of the subsets of $0^+$ states, described above.  

The energies of the $0^+_m$ states in the X(3) model are given in Table II.  Normalizing to the first excited $0^+$ state, $0_2^+$, the energies are described {\sl  exactly}  by 
\begin{equation}
E(0^+_m)=An(n+2), \qquad n=m-1, 
\end{equation}
where $n$ gives the sequencing of $0^+$ states defined such that the first excited $0^+$ state corresponds to $n=1$, 
and $A$ is a scaling factor. 

In the Z(4) model, normalizing to the first excited $0^+$ state, the $0^+_m $ states increase {\sl approximately} as 
\begin{equation}
E(0^+_m)=An(n+2.5), \qquad n=m-1,
\end{equation}
where again, $A$ is a scaling factor dependent on the particular model. 

The energies of $0^+_m$ states in the E(5), Z(5), and X(5) models, normalized to the $2_1^+$ state, are given in Table II.  While these energies are quite different, by normalizing to the first excited $0^+$ state, the models produce exactly identical results, as seen on the right of Table II.  These energies very closely follow the simple formula
\begin{equation}\label{5zero}
E(0^+_m) = An(n+3), \qquad n=m-1.
\end{equation}

Equation (\ref{5zero}) is {\sl not} an exact description of the $0^+$ energies in E(5), Z(5), and X(5); however, it does provide a very accurate approximation.  Through $n$=10, the model $0^+$ energies deviate from the expression given in Eq. (\ref{5zero}) on the order of less than 0.1 $\%$.  

The above empirical results can all be combined into a single, simple formula describing the $0^+_m$ states in any model with an infinite square well potential by 
\begin{equation}\label{zer}
E(0^+_m) = An\left(n+\frac{D+1}{2}\right), \qquad n=m-1
\end{equation}
where $D$ is the number of dimensions and $A$ again depends on the model.    Eq.~(\ref{zer}) is exact only for $D=3$.  As mentioned previously, for other values of $D$ and low values of $\nu$, it is a very accurate approximation. 

\begin{table}
\caption{Energies of $0^+_m$ states in different geometrical models using an infinite square well potential. Columns labelled $2_1^+$ ($0_2^+$) are normalized to the first excited $2^+$ ($0^+$) energies.}
\begin{tabular}{c|cc|cc|cccc}
\hline
$0_m^+$ & X(3) & X(3) & Z(4) & Z(4) & E(5) & Z(5) & X(5) & E(5),Z(5),X(5) \\
\hline
 & $2_1^+$ & $0_2^+$ & $2_1^+$ & $0_2^+$ & $2_1^+$ & $2_1^+$ & $2_1^+$ & $0_2^+$\\
\hline
$0_1^+$ & 0 & 0 &  0 & 0 & 0 & 0 & 0 & 0 \\

$0_2^+$ & 2.87 & 1.0 &  2.95 & 1.0 & 3.03 & 3.91 & 5.65 & 1.0 \\
$0_3^+$ & 7.65 & 2.67 &  7.60 & 2.57 & 7.58 & 9.78 & 14.12 & 2.50 \\
$0_4^+$ & 14.34 & 5.00 &  13.93 & 4.71 & 13.64 & 17.61 & 25.41 & 4.50 \\
$0_5^+$ & 22.95 & 8.00 &  21.95 & 7.43 & 21.22 & 27.39 & 39.53 & 7.00 \\
$0_6^+$ & 33.47 & 11.67 &  31.65 & 10.72 & 30.31 & 39.12 & 56.47 & 10.00 \\
\hline
\end{tabular}
\end{table}

\begin{table*}
\caption{Exact spectra of several Bessel functions $J_\nu$ (labeled by $E_\nu(n)$)
compared to the corresponding 
$n(n+x)$ approximate expressions (labeled by $x$). $J_{3/2}$ occurs in E(5), X(5), Z(5).
$J_{1/2}$ occurs in X(3). $J_1$ occurs in Z(4). $J_0$, $J_1$, $J_2$, $J_3$ 
occur in the pairing case~\cite{pairing}.}

\begin{tabular}{ l | c c | c c | c c | c c | c c | c c | c c }
\hline 
n & $E_0(n)$ & x=1.5 & $E_{1/2}(n)$ & x=2 & $E_1(n)$ &  x=2.5 & $E_{3/2}(n)$ & x=3 & $E_2(n)$ & x=3.5 & $E_{5/2}(n)$ & x=4 & $E_3(n)$ & x=4.5 \\
  
\hline

1 &  1.000 &  1.0 &  1.000 &  1.000 &  1.000 &  1.000 &  1.000 &  1.0 &  1.000 &  1.000 &  1.000 &  1.0 & 1.000 & 1.000\\ 
2 &  2.799 &  2.8 &  2.667 &  2.667 &  2.572 &  2.571 &  2.500 &  2.5 &  2.443 &  2.444 &  2.397 &  2.4  & 2.358 & 2.364\\
3 &  5.398 &  5.4 &  5.000 &  5.000 &  4.715 &  4.714 &  4.499 &  4.5 &  4.329 &  4.333 &  4.192 &  4.2  & 4.077 & 4.091\\
4 &  8.796 &  8.8 &  8.000 &  8.000 &  7.430 &  7.429 &  6.999 &  7.0 &  6.659 &  6.667 &  6.385 &  6.4  & 6.157 & 6.182\\
5 & 12.993 & 13.0 & 11.667 & 11.667 & 10.716 & 10.714 &  9.998 & 10.0 &  9.433 &  9.444 &  8.977 &  9.0  & 8.599 & 8.636\\
6 & 17.990 & 18.0 & 16.000 & 16.000 & 14.574 & 14.571 & 13.497 & 13.5 & 12.651 & 12.667 & 11.968 & 12.0 & 11.403 & 11.455\\
7 & 23.787 & 23.8 & 21.000 & 21.000 & 19.004 & 19.000 & 17.496 & 17.5 & 16.313 & 16.333 & 15.357 & 15.4  & 14.568 & 14.636\\
8 & 30.383 & 30.4 & 26.667 & 26.667 & 24.005 & 24.000 & 21.995 & 22.0 & 20.418 & 20.444 & 19.145 & 19.2  & 18.095 & 18.182\\
9 & 37.779 & 37.8 & 33.000 & 33.000 & 29.577 & 29.571 & 26.993 & 27.0 & 24.967 & 25.000 & 23.332 & 23.4 & 21.983 & 22.091\\
10& 45.974 & 46.0 & 40.000 & 40.000 & 35.721 & 35.714 & 32.492 & 32.5 & 29.960 & 30.000 & 27.918 & 28.0 & 26.233 & 26.364\\

\hline

\end{tabular}
\end{table*}

Equation (4) stems from particular relations between the zeros of the Bessel functions which are involved in the solutions to the infinite square well models.  Given the Bessel function $J_\nu(z)$, with roots $z_s$, $s=1$, 2, 3, \dots we empiricaly observe that the following 
approximate relation holds
\begin{equation}\label{spectr}
E_\nu(n) = {z_n^2 - z_0^2 \over z_1^2-z_0^2} = {n \left( n + \nu +{3\over 2} \right) \over \nu+{5\over 2} },
\end{equation}
where $n=s-1$. 
We call $E_\nu(n)$ the spectrum of the roots of $J_\nu$, since this quantity corresponds to energy spectra in models describing atomic nuclei. 

The relation given in Eq. (5) is exact only in the case $\nu=1/2$, as one can see \cite{private} 
from the expansions of roots of Bessel functions given in \cite{AbrSt} (Eq. 9.5.12).
Numerical results are shown in Table III. 
It is clear that the approximation deteriorates rather slowly with increasing $n$ (while keeping 
$\nu$ constant), while it deteriorates faster with increasing $\nu$ (while keeping $n$ constant).  

For the Bessel function $Y_\nu(z)$, with roots $z_s$, $s=1$, 2, 3, \dots (not used in this paper), we observe that the following 
approximate relation holds
\begin{equation}
E_\nu(n) = {z_n^2 - z_0^2 \over z_1^2-z_0^2} = {n \left( n + \nu +{1\over 2} \right) \over \nu+{3\over 2}},
\end{equation}
where $n=s-1$.
This formula is exact only in the case $\nu=1/2$.  
Similar to the results obtained for the Bessel function $J_{\nu}$($z$) relations, the approximation deteriorates rather slowly with increasing $n$ (while keeping 
$\nu$ constant), while it deteriorates faster with increasing $\nu$ (while keeping $n$ constant).

These results are applicable outside of the models discussed above. As an example, Eq.~(\ref{zer}) applies to a recent model~\cite{pairing} describing the critical point of a pairing vibration to pairing rotation phase transition. Here the energies of $0^+$ states, which span two degrees of freedom, the excitation energies of a particular nucleus and the masses along a series of even-even nuclei, can be described.  In addition, these results would be applicable to hadronic spectra, which have recently been described~\cite{hadron} in terms of roots of Bessel functions. 

We can go further and generalize Eq.~(\ref{5zero}) for a broader range of models.  
The Bohr Hamiltonian can also be solved with potentials in $\beta$ of the form 
$V$ $\sim$ $\beta^{2k}$, giving the so-called  E(5)-$\beta^{2k}$ model~\cite{BonE5} and the X(5)-$\beta^{2k}$ model~\cite{BonX5}, using the $\gamma$ dependence characteristic of E(5) and X(5), respectively.  These models allow for a description of structure between vibrational-like and the infinite square well solutions by increasing the power of $\beta$ in the potential.  For example, in the E(5)-$\beta^{2k}$ case, $\beta^2$ gives the vibrational limit and as the power of $\beta$ goes to infinity, the E(5) solution is reached.   
The evolution of both models is included schematically in Fig. 1(bottom). The predicted $0^+$ energies of the $\beta^{2k}$ models are plotted in Fig. 3(top) normalized to the first $2^+$ state energy.  As evident from Fig. 3(top), with increasing powers of $\beta$ in the potential, the $0^+$ energies evolve gradually towards the infinite square well predictions.  However, again, the E(5) and X(5) related models seemingly give different results.  If instead, we normalize each energy to that of the first excited $0^+$ energy, these models produce exactly identical results for a given $\beta^{2k}$ potential, as seen in Fig. 3(bottom). The normalized $E(0_m^+)$ energies can be reproduced with a generalized version of Eq.~(\ref{5zero}) given by
\begin{equation}\label{genzero}
E(0^+_m) = An(n+x), \qquad n=m-1, 
\end{equation}
where $x$ is some number. The values of $x$ obtained by fitting the first two $0^+$ state energies in each model with Eq.~(\ref{genzero}) are included in Fig. 3. For a harmonic oscillator potential in $\beta$, $x=\infty$, since the bandhead $0^+$ energies increase linearly (i.e., when considering the ratio of energies the term in parenthesis in Eq. (7) disappears).  As the power of $\beta$ in the potential is increased, the value of $x$ decreases, reaching the limiting value of 3 for the infinite square well.  

\begin{figure}
\includegraphics[height=110mm]{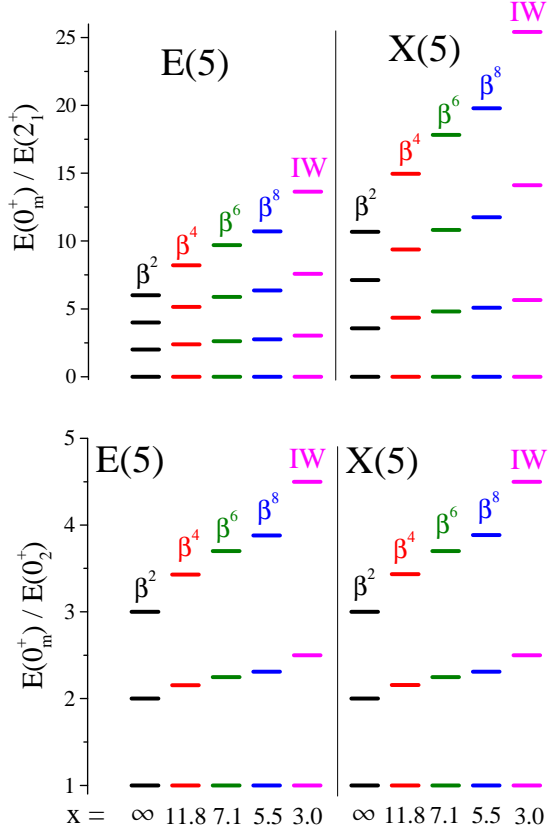}
\caption{(Color online) (Top) Excited $0^+$ state energies in the E(5)-$\beta^{2k}$ and X(5)-$\beta^{2k}$ models, normalized to the $2_1^+$ state energy. (Bottom) Same as top, but with the energies normalized to the first excited $0^+$ state energy.
IW is the infinite square well potential used in the original E(5) and X(5) solutions.}
\end{figure} 

\section{$0^+$ states in the euclidean algebras E($n$)} % 3 

In the solutions of the Bohr Hamiltonian with 
an infinite square well potential in the $\beta$ degree of freedom, the regularities observed for $0^+$ states can be related to the second order Casimir 
operator of E($D$), the Euclidean group in $D$ dimensions. 
In order to see this, 
one can consider in general the Euclidean algebra in $D$ dimensions, E($D$), 
which is the semidirect sum \cite{Wyb} of the algebra T$_D$ of translations 
in $D$ dimensions, generated by the momenta
\begin{equation}\label{eq:e52} 
P_j = -i {\partial \over \partial x_j}, 
\end{equation} 
\noindent and the SO(D) algebra 
of rotations in $D$ dimensions, generated by the angular momenta
\begin{equation}\label{eq:e53} 
 L_{jk} =-i \left(x_j{\partial \over \partial x_k} -x_k {\partial \over
\partial x_j} \right), 
\end{equation}
\noindent symbolically written as E(D) = T$_{\rm D}$ $\oplus_s$ SO(D) \cite{Barut}.  

The generators of E(D) satisfy the commutation relations 
\begin{equation}\label{eq:e54} 
 [P_i, P_j] =0, \qquad [P_i, L_{jk}] = i ( \delta_{ik} P_j - \delta_{ij} P_k),
\end{equation}
\begin{equation}\label{eq:e55} 
 [L_{ij}, L_{kl}]=i (\delta_{ik} L_{jl} +\delta_{jl} L_{ik} 
-\delta_{il} L_{jk} -\delta_{jk} L_{il}).
\end{equation}
\noindent From these commutation relations, the square of the total 
momentum, $P^2$, is a second order Casimir operator of the algebra, while 
the eigenfunctions of this operator satisfy the equation 
\begin{equation}\label{eq:e56} 
 \left( -{1\over r^{D-1}} {\partial \over \partial r} r^{D-1} {\partial \over 
\partial r} + { \omega(\omega+D-2) \over r^2} \right) F(r) = k^2 F(r), 
\end{equation} 
\noindent where on the left hand side of Eq. (\ref{eq:e56}) the eigenvalues of the Casimir operator 
of SO(D), $\omega(\omega+D-2)$ appear \cite{Mosh1555}. 
Using the transformation
\begin{equation}\label{eq:e57} 
 F(r) = r^{(2-D)/2} f(r), 
\end{equation}
\noindent and
\begin{equation}\label{eq:e58}
  \nu= \omega+{D-2\over 2},
\end{equation}
\noindent Eq. (\ref{eq:e56}) can be written as 
\begin{equation}\label{eq:e59} 
 \left( {\partial^2 \over \partial r^2} + {1\over r} {\partial \over \partial 
r} + k^2 - { \nu^2 \over r^2}\right) f(r)  =0, 
\end{equation}
\noindent the eigenfunctions of which are the Bessel functions $f(r) = J_\nu(kr) $
\cite{AbrSt}. 

The ``radial'' equations in the infinite square well models 
E(5) \cite{IacE5}, X(5) \cite{IacX5}, Z(5) \cite{Z5}, Z(4) \cite{Z4}, 
and X(3) \cite{X3} are obtained, 
after the transformation of Eq. (\ref{eq:e57}) has been performed, in the 
form of Eq. (\ref{eq:e59}), with the order $\nu$ summarized in Table I. 

In E(5), Eq. (\ref{eq:e58}) and the corresponding order $\nu$ given in Table I coincide with 
$D=5$ and $\omega =\tau$, where $\tau(\tau+3)$ represents the eigenvalues 
of the second order Casimir operator of SO(5). Thus all states obey Eq. (\ref{eq:e59}). 

In X(5), where again $D=5$, Eq. (\ref{eq:e58}) and the corresponding order $\nu$ given in Table I would agree 
for $J(J+1)/3=\omega(\omega+3)$. This does not hold for any $J$ in general, but it is satisfied for $J=0=\omega$. Thus all $0^+$ bandheads obey Eq. (\ref{eq:e59}). 

In Z(5), where again $D=5$, Eq. (\ref{eq:e58}) and the corresponding order $\nu$ given in Table I would agree
for $n_w=0$ and  
for $J(J+4)/4=\omega(\omega+3)$. Again, this does not hold for any $J$ in general, but it is satisfied for $J=0=\omega$. Thus $0^+$ bandheads with $n_w=0$ obey Eq. (\ref{eq:e59}). 

In the case of Z(4) \cite{Z4}, in which $D=4$, Eq. (\ref{eq:e58}) and the corresponding order $\nu$ given in Table I 
for $n_w=0$ obtain the form $\nu=\omega+1$ and $\nu=J/2+1$, respectively. 
They agree for $J=2\omega$, as already known \cite{Z4}; 
therefore states with any even $J$ and $n_w=0$ obey Eq. (\ref{eq:e59}). 

In X(3), where  $D=3$, Eq. (\ref{eq:e58}) and the corresponding order $\nu$ given in Table I  would agree 
for $J(J+1)/3=\omega(\omega+1)$. Once again, this does not hold for any $J$ in general, but it is satisfied for $J=0=\omega$. Thus $0^+$ bandheads obey Eq. (\ref{eq:e59}). 

The above situation is similar to a partial dynamical symmetry \cite{AlhLev} of Type I
\cite{Lev98}, where some of the states (the $0^+$ states in the present case) 
preserve all the relevant symmetry. 

\section{The IBA Hamiltonian and symmetry triangle} % 4

In order to describe a wider range of structures, it is useful to use a more general collective model than the specific solutions described above.  To this end, we exploit the IBA model, which covers a gamut of structures with an economy of parameters. To do so, we use an IBA Hamiltonian of the form \cite{Volker}
\begin{equation}\label{eq:IBMH}
H(\zeta,\chi) = c \left[ (1-\zeta) \hat n_d -{\zeta\over 4 N_B}
\hat Q^\chi \cdot \hat Q^\chi\right],
\end{equation}
\noindent where $\hat n_d = d^\dagger \cdot \tilde d$, $\hat Q^\chi =
(s^\dagger \tilde d + d^\dagger s) +\chi (d^\dagger \tilde
d)^{(2)},$ $N_B$ is the number of valence bosons, and $c$ is a
scaling factor. The above Hamiltonian contains two parameters,
$\zeta$ and $\chi$, with the parameter $\zeta$ ranging from 0 to
1, and the parameter $\chi$ ranging from 0 to $-\sqrt{7}/2$. The U(5) symmetry is given by $\zeta=0$, any $\chi$, the SU(3) symmetry by $\zeta=1$ and $\chi=-\sqrt{7}/2$, and the O(6) symmetry by $\zeta=1$ and $\chi=0$. 
With this parameterization,  the entire symmetry triangle, shown in Fig.~1, can be described, along with each of the
three dynamical symmetry limits. 
Calculations in this work have been performed with the code IBAR \cite{IBAR,ibar2}, which has recently been
developed to handle large boson numbers.

In Section V, we discuss the fact that, in all three limiting symmetries of the IBA, the energies of certain subsets of $0^+$ states exhibit harmonic behavior in the limit of large boson numbers. 
(This result has also been derived~\cite{largeN1,largeN2} using the coherent state formalism).
In contrast, we will show, in Section VI, that near the critical point, the same subsets
of $0^+$ states exhibit the $n(n+3)$ behavior found in the framework of geometrical models in Section II.
Furthermore, near the critical point these $0^+$ states in the large boson number limit of IBA exhibit certain degeneracies with alternate members 
of the ground state band, calling for further investigation.  

\section{$0^+$ states in the limiting symmetries of the Interacting Boson Model} % 5

We begin an analysis in the IBA framework~\cite{iba} by looking at the three dynamical symmetry limits, 
and analyzing the behavior of the $0^+$ states in the analytic formulae appropriate to each, especially 
in the large $N_B$ limit. Again, we consider a particular subset of $0^+$ states, looking for simple patterns
common to all three symmetries, despite the diversity of the structures they describe.  

In the case of U(5), states are labeled by their quantum numbers $\rm{v}$ and $n_{\Delta}$, where $\rm{v}$ is the seniority and $n_{\Delta}$ is the number of triplets of bosons coupled to angular momentum zero.  There are two classes of $0^+$ states, those with  $n_{\Delta}$ = 0, and those with $n_{\Delta}$ = 1,2,3$\dots$ which are always found degenerate with $3^+$ states. 
In the present work, we consider those states with  $n_{\Delta}$ = 0, that is those $0^+$ states that are not degenerate with $3^+$ states. These states correspond to base states on which major families of levels are built. 
In the U(5) limit, the energies of the $0^+$ states with $n_{\Delta}$=0 are proportional to the number of $d$ bosons, $n_d$, corresponding to their respective phonon number (terms proportional to $n_d^2$ are also allowed, but are omitted in the present consideration); 
thus, the energies increase linearly.   

In the SU(3) limit of the IBA, the position of the $0^+$ bandheads is determined by the second order 
Casimir operator of SU(3).  The eigenvalue expression for $0^+$ states, in terms of the representation labels ($\lambda$, $\mu$), is given by $E$ = $a$[$\lambda^2$ + $\mu^2$ + $\lambda\mu$ + 3($\lambda$ + $\mu$)]. Here, we consider all $0^+$ states.  Taking the $0^+$ state which belongs to the $(2N_B,0)$ irreducible 
representation (irrep) at zero energy, and normalizing to the first excited $0^+$ state, which belongs to the $(2N_B-4,2)$ irrep, 
we find for the lowest $0^+$ states the results shown in Table IV. From Table IV it is clear that at large 
boson numbers $N_B$, we have two states with normalized energy 2, two states with normalized 
energy 3, three states with normalized energy 4, and so on. In other words, for large $N_B$, 
the energies of the $0^+$ states in the SU(3) limit of the IBA grow linearly. 

In the case of O(6), states are labelled by their quantum numbers $\sigma$ and $\tau$.  One set of excited $0^+$ states is found within the multiplet structure of a given $\sigma$ family, has $\tau$ values of 3 or larger and always appears degenerate with $J=6^+,4^+$, and $3^+$ states.  The other set of excited $0^+$ states forms the bandheads of the different $\sigma$ families, has $\tau=0$ and are not degenerate with other states in the spectrum.  We consider only those states with $\tau$=0. In the O(6) limit of the IBA, the positions of the $0^+$ bandheads are determined by the second order Casimir 
operator of SO(6).  The eigenvalue expression for $\tau$ = 0, $J$ = 0 states in terms of the major family quantum number $\sigma$ is $E$ = $a$ $\sigma(\sigma+4)$, with $\sigma=N_B$, $N_B-2$, \dots, 0 or 1. 
Taking the $0^+$ state which belongs to the $(N_B)$ irrep at zero energy, and normalizing to the first excited $0^+$ state belonging to the 
$(N_B-2)$ irrep, we obtain the results shown in Table IV. We observe that $0^+$ bandheads in the O(6) limit 
of the IBA also grow linearly in the limit of large boson numbers $N_B$. 

\begin{table*}
\caption{Irreducible representations (irreps) of SU(3) and O(6) and the corresponding $0^+$ bandhead energies.  In the case of SU(3), energies are normalized to the $0^+$ bandhead with ($\lambda$, $\mu$) = (2$N$-4,2), while in O(6) to the $0^+$ bandhead with $\sigma$ = ($N$-2). $N$ stands for the boson number, $N_B$.}
\begin{tabular}{ c c c c c c | c c }

\hline 
\multicolumn{6}{c|}{SU(3)} & \multicolumn{2}{c}{O(6)}\\

Irrep ($\lambda$,$\mu$) & $E$($0^+$) & Irrep ($\lambda$,$\mu$) & $E$($0^+$) & Irrep ($\lambda$,$\mu$) & $E$($0^+$) & Irrep($\sigma$) & $E$($0^+$) \\

\hline

(2$N$,0) & 0  &   &  &  & & ($N$) & 0   \\
(2$N$-4,2)  & 1  &  &  &   & & ($N$-2) & 1   \\
(2$N$-8,4)  & (4$N$-6)/(2$N$-1)  & (2$N$-6,0)  & (4$N$-3)/($2N$-1) &  &  &
($N$-4) & 2-(2/$N$)  \\
(2$N$-12,6) & (6$N$-15)/(2$N$-1) & (2$N$-10,2) & (6$N$-10)/(2$N$-1) & &  &                ($N$-6) & 3-(3/$N$)\\
(2$N$-16,8) & (8$N$-28)/(2$N$-1) & (2$N$-14,4) & (8$N$-21)/(2$N$-1) & (2$N$-12,0) & (8$N$-18)/(2$N$-1) & ($N$-8) & 4-(8/$N$) \\ 

\hline
\end{tabular}
\end{table*}

For all three IBA dynamical symmetry limits, the energies of these sets of $0^+$ states are given by
\begin{equation}\label{symm}
E = An
\end{equation}
\noindent in the large $N_B$ limit.  Thus, a single simple formula applies to all three dynamical symmetry limits of the IBA despite the fact that each describes very different structures. 
It is interesting that this harmonic behavior, appearing as a general feature 
of IBA spectra, not only at the three limiting symmetries but also in intermediate situations \cite{largeN1,largeN2}, 
is strongly violated near the critical point, as we shall see in the next section.   
 
\section{$0^+$ state energies and degeneracies in the shape coexistence region of the IBA} % 6

It has been recently observed~\cite{letter} that the line describing the degeneracy $E(6_1^+)$ = $E(0_2^+)$  (where $0_2^+$ is the first excited $0^+$ state) in the symmetry 
triangle of the IBA for large $N_B$ ($N_B=250$) falls within the coexistence region of spherical and deformed shapes, slightly to the right 
of the critical line representing the first order phase transition between U(5) and SU(3). 
Similar results are obtained for the $E(10_1^+)$ = $E(0_3^+)$ and $E(14_1^+)$ = $E(0_4^+)$ 
degeneracies.  These degeneracies are interesting not only because they can possibly be associated with underlying symmetries but also because the degeneracy between $E(0_2^+)$ and $E(6_1^+)$ found near the critical point of the IBA is also approximately given by the X(5) critical point model. In what follows, we investigate further the degree to which the IBA predictions near the critical point are related to simple analytic formulas.   

\begin{table}

\caption{Predictions of the IBA (with $N_B=250$, $\chi=-\sqrt{7}/2$, $\zeta=0.473$)
compared to analytic expressions (see text). On the left, excited $0^+$ energies are compared while on the right, energies in the ground state band are compared. Results are normalized to $E(2_1^+) = 1.0$, the numerical factors accompanying $n(n+3)$ and $J(J+2)$ in the column headings reflecting this 
normalization. }

\bigskip

\renewcommand{\arraystretch}{1.3}
\begin{tabular}{ l c c | l c c }

\hline 

& Analytic & IBA & & Analytic & IBA \\

 $n$  & $\;$ $\frac{3n(n+3)}{2}$ $\;$ &  $\;$ $E$($0_m^+$) $\;$ & $J$ & $\;$ $\frac{J(J+2)}{8}$ $\;$ & $\;$ $E(J)$ $\;$  \\

\hline

& & & 2 &  1.00 &  1.00  \\
& & & 4 &  3.00 &  3.05  \\
1 & 6.00 & 6.08 &  6 &  6.00 &  6.08  \\
& & &  8 & 10.00 &  10.00  \\
2 & 15.00 & 14.85 & 10 & 15.00 & 14.73 \\
& & & 12 & 21.00 & 20.23 \\
3 & 27.00 & 27.57 & 14 & 28.00 & 26.43 \\
& & & 16 & 36.00 & 33.30 \\
4 & 42.00 & 42.55 & 18 & 45.00 & 40.81 \\

\hline
\end{tabular}
\end{table}

Along the U(5)-SU(3) leg ($\chi=-\sqrt{7}/2$) a degeneracy between the $6_1^+$ state and the $0_2^+$ state occurs for $\zeta=0.473$ for $N_B=250$.  This occurs very close to, but just beyond, the critical point ($\zeta_{crit}=0.472$ for $N_B$=250) of the phase transition.  Numerical results of the above IBA calculation are given in Table V and compared with simple analytic formulas. In the first three columns, the first four excited $0^+_m$ states
(normalized to the energy of the $2_1^+$ state) obtained in this calculation are compared to the predictions of Eq.~(\ref{5zero}), i.e., to the $n(n+3)$ formula. Very good agreement is obtained up to $n=4$. 
This result is what might be expected given the similarity of the IBA coherent state energy functional at the critical point with an infinite square well potential.   

As mentioned above, successive ground band members with $J>2$ and $J/2$ odd are nearly degenerate with higher lying $0^+$ states.  
Trying to satisfy simultaneously the degeneracies  $E(6_1^+)$ = $E(0_2^+)$ and $E(10_1^+)$ = $E(0_3^+)$ with the $0^+$ states 
obeying Eq. (\ref{5zero}), and the levels of the ground state band obeying 
a general equation of the form $E(J)=AJ(J+y)$,
one obtains $y=2$, i.e., the levels of the ground state band should grow as
\begin{equation}
E(J) = AJ(J+2),
\end{equation}
\noindent where, again, $A$ is some number. In a very crude interpretation, the $J$($J$+2) empirical result can be thought of as the average of the vibrational limit, where the energies go as $J$, and the rotational limit, where the energies go as $J$($J$+1). 

The relevant connection to the $0^+$ expression is given by 
\begin{equation}\label{link}
J(J+2) = 12n(n+3).
\end{equation} 

The predictions of the above mentioned IBA calculation are compared to the $J(J+2)$ predictions,
which are normalized to the energy of the $2_1^+$ state,  
in the right section of Table V. Good agreement is obtained at lower levels, the deviation reaching 10\% at $J=18$.   Also visible in the table are the approximate degeneracies $E(6_1^+)$ = $E(0_2^+)$, $E(10_1^+)$ = $E(0_3^+)$, $E(14_1^+)$ = $E(0_4^+)$, 
$E(18_1^+)$ = $E(0_5^+)$. These degeneracies hold to the 10 percent level for $J=18$.

In summary, IBA $0^+$ states (in the large boson number limit) near the critical point on the U(5)-SU(3) line exhibit the same $n(n+3)$ behavior seen 
in geometrical models involving infinite square well potentials. Furthermore, these $0^+$ states demonstrate approximate 
degeneracies with alternate members of the ground state band, calling for further investigations.  

\section{Conclusions}

Working within the framework of both algebraic and collective models, we have investigated the energies of subsets of excited $0^+$ states, pointing out regularities within and similarities between the two different approaches.  For models employing an infinite square well potential in the $\beta$ degree of freedom, a single formula is derived for a subset of excited $0^+$ state energies, dependent only on the number of dimensions of the model. 
The same regular behavior for $0^+$ states has been found in IBA calculations (in the large boson number limit) near the critical point of the first order phase transition between U(5) and SU(3), despite the fact that in all three limiting symmetries of the IBA (in the large boson number limit)
the same $0^+$ states exhibit a harmonic behavior. Furthermore, these successive $0+$ states near the critical point exhibit 
degeneracies with alternate yrast states, analogous to the near-degeneracy that occurs between the first $6^+$ state and the first excited $0^+$ state in X(5), calling for further investigations. Finally, the observed regularities in $0^+$ energies are discussed in terms of the underlying group theoretical framework of the different models.  

\section{ACKNOWLEGDEMENTS}

The authors are thankful to M. A. Caprio  for bringing \cite{private} to our attention 
the relevance of the asymptotic formula for the zeros of the Bessel functions 
and for useful discussions. 
This work was supported by U.S. DOE Grant No. DE-FG02-91ER-40609 and 
by the DOE Office of Nuclear Physics under Contract DE-AC02-06CH11357.


\begin{thebibliography}{99}

\bibitem{Oothoudt}
M. A. Oothoudt and N. M. Hintz, Nucl. Phys. A {\bf 213}, 221 (1973).

\bibitem{Lesher}
S. R. Lesher, A. Aprahamian, L. Trache, A. Oros-Peusquens, S. Deyliz, A. Gollwitzer, R. Hertenberger, B. D. Valnion, 
and G. Graw, Phys. Rev. C {\bf 66} , 051305(R) (2002).

\bibitem{des} D.A. Meyer {\it et al.,} Phys. Lett. B {\bf 638}, 44 (2006).

\bibitem{Asai}
M. Asai, T. Sekine, A. Osa, M. Koizumi, Y. Kojima, M. Shibata, H. Yamamoto, and K. Kawade,  
Phys. Rev. C {\bf 56}, 3045 (1997).

\bibitem{iba}
F. Iachello and A. Arima, {\it The Interacting Boson Model} (Cambridge 
University Press, Cambridge, 1987). 

\bibitem{gilmore}
D. H. Feng, R. Gilmore, and S. R. Deans, Phys. Rev. C {\bf 23}, 1254 (1981). 

\bibitem{GK}
J. N. Ginocchio and M. W. Kirson, Phys. Rev. Lett. {\bf 44}, 1744 (1980).

\bibitem{DSI}
A. E. L. Dieperink, O. Scholten, and F. Iachello, Phys. Rev. Lett. {\bf 44}, 1747 (1980). 

\bibitem{ricktri}
R. F. Casten, {\it Nuclear Structure from a Simple Perspective}
(Oxford University Press, Oxford, 1990).

\bibitem{IZC}
F. Iachello, N. V. Zamfir, and R. F. Casten, Phys. Rev. Lett. {\bf 81}, 1191 (1998).

\bibitem{Jolie89}
J. Jolie, P. Cejnar, R. F. Casten, S. Heinze, A. Linnemann, and V. Werner, Phys. Rev. Lett. {\bf 89}, 
182502 (2002). 

\bibitem{IacE5} F. Iachello, Phys. Rev. Lett.  {\bf 85}, 3580 (2000). 

\bibitem{IacX5} F. Iachello, Phys. Rev. Lett. {\bf 87},  052502 (2001). 

\bibitem{bohr} A. Bohr, Mat. Fys. Medd. K. Dan. Vidensk. Selsk. {\bf 26}, no. 14 (1952). 

\bibitem{CZE5}
R. F. Casten and N. V. Zamfir, Phys. Rev. Lett. {\bf 85}, 3584 (2000). 


\bibitem{CZX5} % 152-Sm
R. F. Casten and N. V. Zamfir, Phys. Rev. Lett. {\bf 87}, 052503 (2001). 

\bibitem{Kruecken} % 150-Nd
R. Kr\"{u}cken, {\it et al.}, Phys. Rev. Lett. {\bf 88}, 232501 (2002).

\bibitem{tonev} D. Tonev, A. Dewald, T. Klug, P. Petkov, J. Jolie, A. Fitzler, O. M\"{o}ller, S. Heinze, P. von Brentano, and R.F. Casten,  Phys. Rev. C {\bf 69}, 034334 (2004).

\bibitem{frank} A. Frank, C.E. Alonso, and J. M. Arias, Phys. Rev. C {\bf 65}, 014301 (2001).

\bibitem{JPGreview}
R. F. Casten and E. A. McCutchan, J. Phys. G: Nucl. Part. Phys. {\bf 34}, R285 (2007). 

\bibitem{Rowe735}
D. J. Rowe, Nucl. Phys. A {\bf 735}, 372 (2004). 

\bibitem{Rowe753}
D. J. Rowe and P. S. Turner, Nucl. Phys. A {\bf 753}, 94 (2005).

\bibitem{RWC}
D. J. Rowe, T. A. Welsh, and M. A. Caprio, Phys. Rev. C {\bf 79}, 054304 (2009). 

\bibitem{Thiamova1}
D. J. Rowe and G. Thiamova, Nucl. Phys. A {\bf 760}, 59 (2005). 

\bibitem{Thiamova2}
G. Thiamova and D. J. Rowe, Czech. J. Phys. {\bf 55}, 957 (2005). 

\bibitem{ramos} J.E. Garc\'{i}a-Ramos and J.M. Arias, Phys. Rev. C {\bf 77}, 054307 (2008).  

\bibitem{mapping} E.A. McCutchan, N.V. Zamfir, and R.F. Casten, Phys. Rev. C {\bf 69}, 064306 (2004). 

\bibitem{cejnar} P. Cejnar and J. Jolie, Phys. Rev. E {\bf 61}, 6237 (2000). 

\bibitem{pgarrett}
P. E. Garrett, J. Phys. G: Nucl. Part. Phys. {\bf 27}, R1 (2001). 

\bibitem{wchou} W.-T. Chou, Gh. Cata-Danil, N.V. Zamfir, R.F. Casten, and N. Pietralla, Phys. Rev. C {\bf 64}, 057301 (2001).

\bibitem{nds} http://www.nndc.bnl.gov/ensdf/


\bibitem{map} D. Bonatsos, E.A. McCutchan, and R.F. Casten, Phys. Rev. Lett. {\bf 101}, 022501 (2008).

\bibitem{letter} D. Bonatsos, E.A. McCutchan, R.F. Casten, and R.J. Casperson, Phys. Rev. Lett. {\bf 100}, 142501 (2008). 

\bibitem{Z5}
D. Bonatsos, D. Lenis, D. Petrellis, and P. A. Terziev, Phys. Lett. B {\bf  588}, 172 (2004). 

\bibitem{Z4}
D. Bonatsos, D. Lenis, D. Petrellis, P. A. Terziev, and I. Yigitoglu, 
Phys. Lett. B {\bf  621}, 102 (2005). 

\bibitem{X3}
D. Bonatsos, D. Lenis, D. Petrellis, P. A. Terziev, and I. Yigitoglu, 
Phys. Lett. B {\bf 632}, 238 (2006).  

\bibitem{BM}
A. Bohr and B. R. Mottelson, {\it Nuclear Structure, Vol. II: Nuclear Deformations} (Benjamin,
New York, 1975). 

\bibitem{private}
M. A. Caprio, private communication (2008). 

\bibitem{AbrSt}
M. Abramowitz and I. A. Stegun, {\it Handbook of Mathematical Functions} 
(Dover, New York, 1965). 

\bibitem{pairing}
R. M. Clark, A. O. Macchiavelli, L. Fortunato, and R. Kr\"ucken, Phys. Rev. Lett. 
{\bf 96}, 032501 (2006). 

\bibitem{hadron} G.F. de T\'{e}ramond and S.J. Brodsky, Phys. Rev. Lett. {\bf 94}, 201601 (2005).

\bibitem{BonE5}
D. Bonatsos, D. Lenis, N. Minkov, P. P. Raychev, and P. A. Terziev, Phys. Rev. C {\bf 69}, 044316 (2004). 

\bibitem{BonX5}
D. Bonatsos, D. Lenis, N. Minkov, P. P. Raychev, and P. A. Terziev, 
Phys. Rev. C {\bf 69}, 014302 (2004).

\bibitem{Wyb}
B. G. Wybourne, {\it Classical Groups for Physicists} (Wiley, New York, 1974). 

\bibitem{Barut}
A. O. Barut and R. Raczka, {\it Theory of Group Representations and 
Applications} (World Scientific, Singapore, 1986). 

\bibitem{Mosh1555}
M. Moshinsky, J. Math. Phys.  {\bf 25}, 1555 (1984). 

\bibitem{AlhLev}
Y. Alhassid and A. Leviatan, J. Phys. A: Math. Gen. {\bf 25}, L1265 (1992). 
 
\bibitem{Lev98}
A. Leviatan, Phys. Rev. Lett. {\bf 98}, 242502 (2007). 

\bibitem{Volker}
V. Werner, N. Pietralla, P. von Brentano, R. F. Casten, and R.V. Jolos, Phys. Rev. C {\bf 61}, 021301(R) (2000).

\bibitem{IBAR}
R. J. Casperson, IBAR code (unpublished). 

\bibitem{ibar2} E. Williams, R.J. Casperson, and V. Werner, Phys. Rev. C {\bf 77}, 061302(R) (2008). 

\bibitem{largeN1} A. Leviatan, Ann. Phys. (NY) {\bf 179}, 201 (1987).

\bibitem{largeN2} J.E. Garcia-Ramos, C.E. Alonso, J.M. Arias, P. Van Isacker, and A. Vitturi,
Nucl. Phys. A {\bf 637}, 529 (1998).  

\end{thebibliography}
\end{document}